\documentclass[twocolumn,amsmath,amssymb]{article}
\usepackage{graphicx}
\usepackage{epstopdf}
\advance\topmargin-1in\advance\oddsidemargin-1in
\evensidemargin\oddsidemargin

\makeatletter
\def\@normalsize{\@setsize\normalsize{12pt}\xpt\@xpt
\abovedisplayskip 10pt plus2pt minus5pt\belowdisplayskip \abovedisplayskip
\abovedisplayshortskip \z@ plus3pt\belowdisplayshortskip 6pt plus3pt
minus3pt\let\@listi\@listI}

\def\section{\@startsection {section}{1}{\z@}{20pt plus 2pt minus 2pt}
{8pt plus 2pt minus 2pt}{\centering\normalsize\sc
\edef\@svsec{\thesection.\ }}}
\def\thesection{\Roman{section}}

\def\subsection{\@startsection {subsection}{2}{\z@}{16pt plus 2pt minus 2pt}
{6pt plus 2pt minus 2pt}{\normalsize\sl
\edef\@svsec{\thesubsection.\ }}}
\def\thesubsection{\Alph{subsection}}

\long\def\@makecaption#1#2{
\vskip10pt\begin{center} #1 #2 \end{center}\par\vskip 1pt}
\def\fnum@figure{\raggedright{\footnotesize Fig. \thefigure }.%
\footnotesize}
\def\fnum@table{\footnotesize TABLE \thetable\\\footnotesize\sc}
\def\thetable{\Roman{table}}

\makeatother

\begin{document}
\newtheorem{definition}{Definition}[section]
\date{}

\title{\Large\textbf{Low-overhead pieceable fault-tolerant construction of logical controlled-phase circuit for degenerate quantum code}}	


\author{Chen Lin \and GuoWu Yang}

\maketitle
\thispagestyle{empty}

{\small\bf Abstract---
We designed an search algorithm in order to find a non-transversal but fault-tolerant construction of a logical controlled-phase gate for general [[n,1,d]] degenerate quantum code. Then we give an example to illustrate our algorithm for a quantum code called bare [[7, 1, 3]] code. This code is obtained under certain search criteria, and it possesses a simpler flag-assisted fault-tolerant syndrome measurement circuit under a standard depolarizing error model. Since a syndrome extraction circuit requiring fewer ancillary qubit resources would facilitate the realization of large-scale quantum computations, such as concatenated high level elementary logical gate circuits when aim to achieve lower logical error rates, we follow our search scheme and find a 3-pieceable fault-tolerant logical CZ circuit on this code. Numerical simulations are also performed to further analyze the logical error rate of our circuit.}

\section{Introduction}

Practical quantum computation may be coming soon\cite{preskill2018quantum,neill2018blueprint}. Consequently, the prospect of large-scale quantum computers has generated interest, as they can solve certain problems exponentially faster than can computers using the best known classical algorithms\cite{shor1999polynomial}. However, given the limitations of current quantum computers,  called noisy intermediate-scale quantum (NISQ) computers, high-fidelity quantum circuits for the compiling of certain quantum algorithms always require very large quantum resources\cite{campbell2017roads,gottesman2013fault}. More specifically, assume that the physical components, such as physical quantum gate,ancillary preparation and measurement, could fail with probability at most \emph{p} based on a certain experimental environment. When \emph{p} is below a certain constant threshold\cite{aliferis2005quantum}, it has been proved that, any quantum circuit consisted of $n$ physical elementary components can be simulated fault-tolerantly with a cost up to $O(poly(log(\frac{n}{\varepsilon})n)$, where $\varepsilon$ is the desired accuracy\cite{knill1996accuracy,nielsen2000quantum}.

Conversely, instead of combining physical qubits through the use of error-correcting codes and providing fault-tolerant logical operations\cite{gottesman2010introduction}, if we directly implement  a computation task on a given quantum system, the noisy elementary components will cause uncorrelated faults with high probability and strongly propagate throughout the various quantum systems that are being coupled for the use of quantum computation. Moreover, from the perspective of quantum circuit synthesis, the construction of a set of universal fault-tolerant logical quantum gates is quite significant for quantum computations. The terminology "universal" can be described as any quantum operation can be composed of a set of finite quantum gates. However, for most fault-tolerant schemes, additional resources are always required to provide universal quantum computation\cite{bravyi2005universal,fowler2013surface}.

Since the cost of implementing logical gate operations is usually considered to be a major figure of merit of fault-tolerant quantum computations, many studies have focused on designing efficient fault-tolerant protocols. One of the most widely used and simplest such schemes is transversal operations. A logical quantum operation with transversal construction is automatically fault-tolerant, as each qubit acted upon by this logical circuit can interact with no more than one qubit per each code block. For example, quantum error correction code of a stabilizer form, such as the smallest distance-3 code, called a 5-qubit quantum code, is denoted as [[5,1,3]] code. This code has several important transversal logical gates such as the Hadamard gate and phase gate. Therefore, it is worth finding a universal transversal logical gate set because this could effectively lower resource requirements for fault-tolerant quantum computations\cite{metodi2005general,oskin2002practical}. Unfortunately, no quantum code can implement a universal logical gate set on the encoded data\cite{zeng2011transversality}. This fact encourages the development of other methods to circumvent this limitation. Another common approach is called magic state distillation\cite{bravyi2005universal,bravyi2012magic}; although this method can promote available transversal circuits to a universal set of logical circuits through gate teleportation, it actually requires large overhead in terms of time and number of qubits. Concatenating techniques have also been used to construct universal fault-tolerant logical operations and further improve the accuracy threshold but at the cost of exponential qubit resource overhead\cite{jochym2014using}.

Recently, a novel paradigm for fault-tolerant logical circuit design has been proposed by Yoder\cite{yoder2016universal}, which focuses on creating non-transversal yet fault-tolerant logical quantum circuits based on stabilizer code. By dividing the non-fault-tolerant logical circuit into several pieces and inserting a procedure called constant stabilizer error correction between these pieces to correct contagious errors, they developed an adaptive decoding procedure to obtain a fault-tolerant variant of this logical circuit. However, for degenerate code, there are no sufficient conditions for which there exists a good decoding algorithm that makes its round-robin logical circuit fault tolerant. On the other hand, the implementation of this kind of fault tolerant logical circuit usually requires substantial ancillary qubit resources due to several rounds of intermediate error correction , which makes it more resource intensive when serving as the elementary logical component in high level concatenation circuit. Various methods have been proposed to make the syndrome measurement fault tolerant; The ancillary qubit consumption of Shor's method can be determined based on the weight of parity checks\cite{shor1996fault}, the number of qubits in a code block\cite{steane1996error}, or the resource requirements of the fault-tolerant preparation of a logical Bell pair\cite{knill1996accuracy}. Some effort is being invested into reducing the resource requirements associated with such fault-tolerant syndrome measurement schemes such as subsystem codes\cite{bacon2006operator,aliferis2007subsystem}. Lately, it has been observed that, for certain types of degenerate quantum codes\cite{li2017fault}, fault-tolerant syndrome measurement circuits can be realized by using a single ancillary qubit per stabilizer, plus an extra flag qubit \cite{chao2018fault} to signal errors which propagated from ancillary qubit, also called hook error.

In this paper, we first propose a scheme for realizing an ancillary less consumed level-1 fault tolerant logical CZ circuit. Inspired by the round-robin logical circuit construction, we propose a heuristic search algorithm for which to find a decoder, as a classical post-processing procedure for the measurement results collected from the implementation of the circuit, to realize the corresponding CZ circuit on general degenerate stabilizer code [[n,1,d]]. Secondly, for general pieceable fault tolerant logical circuit, we observed that idle qubit error may disturb the intermediate stabilizer measurement, we handle this problem by proposing a non-uniform concatenation method, which is used to further depress the possibility of error propagation during the implementation of the circuit.

To further illustrate our methods, we choose a code that consumes fewer ancillary resources called bare-[[7,1,3]] code\cite{li2017fault} as an example and use our algorithm to construct a 3-pieceable fault-tolerant encoded CZ circuit. Our algorithm guarantees that the propagation through the syndrome measurement circuit is signaled and that a “good” decoder is used; it has also been noted that the logical error rate should follow a power law scaling\cite{fowler2012surface}. Given Monte Carlo error sampling method\cite{li2017fault}, we use the LIQUi$\vert \rangle$ quantum simulation platform\cite{wecker2014liqui} to obtain the logical error rates for our fault-tolerant logical CZ circuit at the first level of encoding, and we verified that, under our decoding procedure, the circuit's error rate goes as the square of the physical error rate, i.e., $\epsilon_{L} \propto \epsilon_{phys}^{2}$. Meanwhile, by uniform concatenate the idle qubit, the logical error rate will be lower than the non concatenated one. We also report the level-1 logical error rate for another pieceable fault-tolerant logical CZ circuit\cite{yoder2016universal} based on a well-studied distance-3 quantum error correcting code, the 5-qubit code, for comparison.

The paper is organized as follows. Section II introduces the preliminaries of quantum error correction (QECC) and fault-tolerant circuit design, and it details the pieceable fault-tolerant circuit scheme and other technique used in this paper. Section III describes our algorithm in terms of fault-tolerant CZ circuit design for general [[n,1,d]] stabilizer code. Section IV  introduces our 3-pieceable fault-tolerant logical CZ circuit based on bare-[[7,1,3]] code, and it gives a further discussion on an adjusted concatenation scheme for further promotion of accuracy of the pieceable logical circuit. Section V presents the results of the resource comparison and Monte Carlo simulation results to compare their logical error rates.

\section{Codes,logical operations and pieceable fault tolerant construction}
This section reviews definitions and preliminary results about QECC and pieceable fault-tolerance circuit design schemes. Throughout this paper, we only consider stabilizer code, which is defined by a set of Pauli operators, also referred to as the stabilizer formalism developed by\cite{gottesman2010introduction}:
\begin{definition}(n-qubit Pauli Group)
The n-qubit Pauli group $G_{n}$ consists of the tensor products of a $2 \times 2$ identity matrix and single-qubit Pauli operators that can be described by the following set:
\begin{equation}
G_{n} = \{ \lambda\sigma_{1}\otimes \sigma_{2} \otimes ... \otimes \sigma_{n}: \sigma_{i} \in \{ I,X,Y,Z\}\}
\end{equation}
\end{definition}
where $\lambda \in \{ \pm 1,\pm i \}$.

Meanwhile, we present the concept of the \emph{support} of a Pauli operator. A support is defined as a subset of $[n] := \{ 1,2,...,n \}$, given an operator $p \in G_{n}$, where we denote by the symbol \emph{supp(p)}  the set of all $i \in [n]$ such that p acting on the i-th qubit is not identity, and the \emph{weight} of p equals the size $\vert supp(p) \vert$ of the support. 

Next, we describe the stabilizer formalism. A \emph{stabilizer} S is an abelian subgroup belonging to $G_{n}$ that does not contain -I. Usually, we define a stabilizer group through a set of generators; because the operators in this group are Hermitian and mutually commuting, they can be diagonalized simultaneously. Thus, we can intuitively define the stabilizer code by the following statement:
\begin{definition}(n-qubit Stabilizer Code)
Given a stabilizer $S \in G_{n}$, an n-qubit stabilizer code S(Q) is the joint eigenspace of S belonging to the Hilbert space $(C^{2})^{\otimes n}$ and can be defined by the following set:
\begin{equation}
S(Q) = \{ \vert \psi \rangle \vert g\vert \psi \rangle = \vert \psi \rangle, \forall g \in S \}
\end{equation}
\end{definition}
without loss of generality, we can assume that S(Q) is of dimension $2^{n-m}$, and we have [[n,k,d]] quantum code, where $k = n-m$ is the number of logical qubits and d is the minimum weight of the Pauli operators that can map the code space onto itself. From the fundamentals of quantum coding theory, we can say that the stabilizer S has m generators, denoted as $S = <g_{1},...,g_{m}>$, and the code S(Q) can correct Pauli errors of weight $t < \lfloor \frac{d-1}{2} \rfloor$.

With the definition of code stated above, it is necessary to find corresponding logical operations to realize a given computation task. The Pauli subgroup C(S) called the \emph{centralizer} of S, defined to be a set of elements in $G_{n}$ that commute with all elements in S. We can first choose operators $\bar{Z}_{1},...,\bar{Z}_{k}$ and $\bar{X}_{1},...,\bar{X}_{k}$ in C(S) that are independent of the generators of S and satisfy the communication conditions $\bar{X}_{i}\bar{Z}_{j} = (-1)^{\delta_{ij}}\bar{Z}_{j}\bar{X}_{i}$. For convenience, we also refer to these operators as logical Pauli operators; therefore, it is not difficult to say that the encoded $\vert \bar{0} \rangle$ can be uniquely represented by the joint +1 eigenspace of the following set of n operators: $<g_{1},...,g_{m},\bar{Z}_{1},...,\bar{Z}_{k}>$. Therefore, we can give other basic encoded states by applying the corresponding logical Pauli X operators to $\vert \bar{0} \rangle$ such that $ \vert x_{1}...x_{k} \rangle_{L} = \prod_{i = 1}^{k} \bar{X}_{i}^{x_{i}}\vert \bar{0} \rangle$, where $x_{i} \in \{ 0,1 \}.$

In addition to the method of protecting information via active error detection and correction when it is transmitted and stored, how to perform operations on an encoded state without losing the code’s protection against errors and how to perform error correction safely when the gates used are themselves noisy are highly important topics. For the basics on fault-tolerant quantum computation, we refer the reader to\cite{gottesman2010introduction,nielsen2000quantum}. We next introduce the following technical details, which can be used to construct a fault-tolerant logical CZ circuit on any stabilizer code, called pieceable fault-tolerant round-robin construction.

In contrast to transversal design schemes, non-transversal circuits can still provide fault tolerance if we partition the circuit into several fault-tolerant circuit pieces and insert the error correction between these pieces. We can then make the entire circuit fault tolerant; this scheme is also called pieceable fault tolerance\cite{yoder2016universal}. It would be convenient if could we intuitively imagine that a certain logical circuit has the following decomposition:

\begin{equation}
C = C_{r}\cdot C_{r-1}...C_{1}
\end{equation}
where r is an integer. We can obtain a fault-tolerant variant of the circuit C if each $C_{i}$ is carefully designed such that propagating errors can be avoided. We now give the modified  circuit C, denoted as $\widetilde{C}$, which can be fault tolerant.

\begin{equation}
\widetilde{C} = EC_{r}\cdot C_{r} \cdot EC_{r-1}\cdot C_{r-1}...EC_{1}\cdot C_{1}
\end{equation}
Actually, by performing the corrective actions after each $C_{i}$ on the encoded data, we obtain several fault-tolerant gadgets $EC_{i}\cdot C_{i}$(i = 1,...,r). For the remaining of this paper,we only consider the 1-fault tolerant circuit design. Next we note some essential concepts related to the this fault tolerant design scheme, first we give the definition of the contagious error; it is a type of Pauli error operator that can be described by the following statement: 

\begin{equation}
\mathcal{E}_{C} = \{ E \in G_{n}: \exists i \quad s.t. [E,C_{i}] \neq 0\}
\end{equation}
However, only contagious errors occurring in $C_{i}$ should been corrected during the intermediate actions $EC_{i}$. During the procedure for $EC_{i}$, we can simultaneously correct contagious errors, infer  possible non-contagious error operators applied to the encoded data during $C_{i}$, and provide the information about the location of non-contagious errors to the final error correction procedure. The new tool for the detection of contagious errors should also be introduced:

\begin{equation}
S_{C} = \{ s \in S: \forall i, [s,C_{i}] = 0 \}
\end{equation}
We can intuitively observe that only contagious errors in a given decomposition of circuit will propagate. For example, if we consider a physical CZ gate as a subcircuit within the decomposition of C, then the input single Pauli X error on either the control qubit or the target qubit will propagate to form a two-qubit error XZ. Based on the above definition, we say that the single Pauli X is a contagious error corresponding to this CZ gate, and the Z-type Pauli errors are non-contagious.

Besides the above mentioned fault tolerant circuit structure, by thinking about the structural details of given nontransversal gate circuit, we can apply appreciate concatenation scheme so as to reduce the overhead or promote the stability for certain component in the circuit\cite{nikahd2017nonuniform,jochym2014using}. More specially, qubits in a single concatenated quantum error correcting code block are re-encoded for further protection. This provides a means to reduce the error rate in a double-exponential manner as the number of concatenation levels increase.

\section{The pieceable FT logical CZ circuit search algorithm}
Having reviewed the method used by the pieceable fault tolerance scheme, we now consider how to design a pieceable fault-tolerant variant of the logical CZ circuit with the round-robin construction\cite{yoder2016universal} for any degenerate [[n,1,d]] stabilizer code while simultaneously showing that our circuit can be divided into only a few circuit pieces.

Actually, for fault tolerance under depolarizing error model, it suffices to only consider gate faults, including the idle time step, which can be treated as an identity quantum gate at that time step; A faulty gate can be seen as an implementation of the ideal gate followed by one of the non-trivial Pauli errors. Therefore, by using an adaptive stabilizer error correction, i.e., each piece applies X-error correction to maintain fault tolerance but refrains from applying Z-error correction until the entire CZ gate circuit is implemented, we propose a heuristic method for searching for such a scheme for the circuit $\Gamma(Z,Z) = CZ_{Logical}$ with the minimum number of immediate constant stabilizer error correction procedures. The following statements gives a general description of our search scheme:
\begin{table}[tb]
\caption{The minimal pieceable FT logical CZ circuit search algorithm}
\begin{minipage}{8cm}
\label{tab:table1}       
\begin{tabular}{l}
\hline\hline\noalign{\smallskip}
\textbf{Input:}Two code blocks A and B encoded by \\
\qquad \quad   a [[n,1,d]] stabilizer code S(Q).\\
\qquad \quad   Round-robin logical CZ circuit: \\
\qquad \quad   $\Gamma(Z,Z) = \prod_{j_{A},k_{B} \in supp(\bar{Z})} CZ(j_{A},k_{B})$,\\
\qquad \quad   based on S(Q)\\
\textbf{Output:}The pieceable FT variant of circuit $\Gamma(Z,Z)$ \\
\qquad \qquad   with minimal number of pieces under \\
\qquad \qquad   some application order\\
(1) For each partition chosen from a partition set with \\
    quantity $C_{m-1}^{r}$.\\
(2) For each application order chosen from a order set\\
    with quantity $m!$, obtain a given r-piece\\
    variant of $\Gamma(Z,Z)$\\
(3) For each $C_{i}$ in $\Gamma(Z,Z)$, compute the combined\\
    syndrome of all possible t-qubit gate errors.\\
(4) If $C_{i}$ possesses two errors $E_{i}^{'}$ and $E_{i}^{''}$ such that\\
    $E_{i}^{'}(Z)E_{i}^{''}(Z) = \in C(S)/S$, then break to step 2, \\
    choose next application order and repeat steps 3-4.\\
(5) If  condition (4) is false, then we obtain the desired\\
    output.\\
(6) If under all application orders we cannot make \\
    the r-piece variant of $\Gamma(Z,Z)$ be an FT circuit, then \\
    go back to step 1 and set the piece number to r+1.\\
\noalign{\smallskip}\hline\hline
\end{tabular}
\end{minipage}
\end{table}
(1) For any stabilizer code S(Q) that encodes one data qubit in a code block, denoted as [[n,1,d]], we first select its logical Pauli Z operator, denoted as $\bar{Z}$, and apply a local Clifford transformation\cite{zeng2011transversality} to obtain the equivalent form of the operator $\bar{Z}$, which has weight d and consists of only positive-signed Pauli Z's and I's. The stabilizer S will also change due to the conjugation of these local Clifford operators.

(2) Now, we construct $\Gamma(Z,Z)$, which can be implemented by applying the physical CZ gate to those active qubits, i.e., qubits whose index is in supp($\bar{Z}$) for both control code block A and target code block B. It has been proved that these gates will form a round-robin logical CZ circuit on S(Q), written as $\Gamma(Z,Z) = \prod_{j_{A},k_{B} \in supp(\bar{Z})}CZ(j_{A},k_{B})$.

(3) Next, we search for a pieceable fault-tolerant variant of $\Gamma(Z,Z)$. Generally we can assume that all m CZ gates in the above product are divided into r pieces at the present time, where $2 < r < m$ and $m = \vert supp(\bar{Z}) \vert^{2}$. Furthermore, the number of different ways to divide $\Gamma(Z,Z)$ into r pieces is $C_{m-1}^{r}$. Then, for each partition, whether $\Gamma(Z,Z)$ is an r-piece fault-tolerant circuit can be determined by the following subroutines:

(a) It has been noted that all these CZ gates in $\Gamma(Z,Z)$ are mutually commute; therefore, $\Gamma(Z,Z)$ can be implemented with $m!$ different application orders, Then, for each order, whether it is an fault-tolerant circuit can be determined by the following procedure:

(i) For each circuit piece $C_{i}(i = 1,...,r)$, we first consider all single t-qubit Pauli gate errors occurring in each piece, denoted as $E_{i}$, where $t \in [1,\lfloor \frac{d-1}{2} \rfloor]$; gather their constant stabilizer syndromes through $EC_{i}$ and final non-contagious error syndromes through $EC_{non-contagious}$; and finally make a correspondence between the combined syndrome vector and these errors.

(ii) Meanwhile, all  the single errors $E_{i}$ will be partially corrected, and the non-contagious parts $E_{i}(Z)$ will remain until the final non-constant error correction procedure. We then propose a condition about whether there exist two errors $E_{i}^{'}, E_{i}^{''} \in E_{i}$ such that the combined syndrome of the two errors is the same but where $E_{i}^{'}(Z)E_{i}^{''}(Z) \in C(S)/S$.

(iii) If for all r circuit pieces there exists a $C_{i}$ such that the condition in (ii) is true, then we can go back to (a), choose the next application order, and continue with procedures (i)-(iii).

(4) If during step 3 we cannot find an fault-tolerant partition scheme for the logical CZ circuit, then we increase the number of pieces to r+1 and continue the above search procedure. Because $\Gamma(Z,Z)$ is composed of a finite number of CZ gates, the search procedure ultimately stops after all these partitioning schemes are applied. We summarize our search algorithm in Table~\ref{tab:table1}. 

\section{Pieceably fault-tolerant CZ on bare [[7,1,3]] code}
\begin{figure}[htb]
\centering\includegraphics[width=0.9\linewidth]{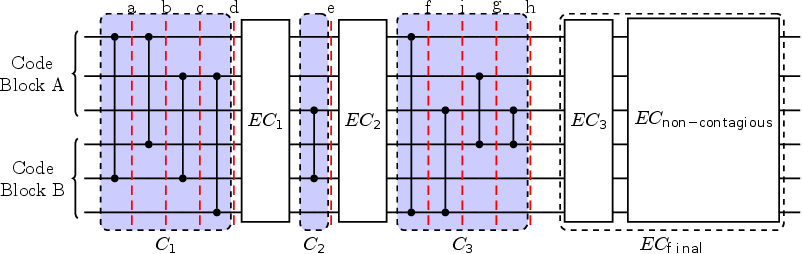}
\caption{\label{fig:figure1}Pieceably fault-tolerant implementation of round-robin logical CZ circuit between two bare [[7,1,3]] code blocks, each with three active qubits, i.e., qubits 0, 1, and 4. Idle code qubits are omitted. We label the control code block as A and the target code block as B. With our search scheme, we propose that this round-robin circuit is a 3-pieceable fault-tolerant circuit, more specifically, $\Gamma(Z,Z) = C_{3}\cdot C_{2}\cdot C_{1}$, the intermediate error correction process applied after $C_{i}(i = 1,2,3)$, during which only the constant stabilizers are measured to correct the Pauli X error. For the non-contagious error correction $EC_{non-contagious}$, we correct non-contagious Pauli Z errors by utilizing the error syndrome information from $EC_{i}(i = 1,2,3)$. Two copies of the canonical error correction are also applied after $EC_{non-contagious}$  to check the possible error caused by the hook error on idle qubits; the dashed box denotes $EC_{final}$. Finally, we take the  circuits illustrated above as an entire circuit and denote it as $\widetilde{CZ}_{logical} = EC_{final}\cdot C_{3}\cdot EC_{2}\cdot C_{2}\cdot EC_{1}\cdot C_{1}$.}
\end{figure}

Motivated by the property of bare [[7,1,3]] code that implies that there exists an optimized configuration of ancillary resources for fault-tolerant syndrome measurement, we develop a pieceably fault-tolerant logical CZ circuit for this code with simpler syndrome extraction circuit. The generators of the stabilizer \emph{S} and logical Pauli operators of  this 7-qubit code\cite{li2017fault} are shown in Table~\ref{tab:table2}
\begin{table}[tb]
\caption{Generators of Stabilizer and Logical Pauli operators for bare [[7,1,3]] code}
\label{tab:table2}
\begin{tabular}{l|l}
\hline \hline \noalign{\smallskip}
\textrm{Stabilizer generators} & \textrm{Logical Pauli Operator} \\
\hline
$X_{0}X_{4}$ &  \\
$X_{1}X_{4}$ &  \\
$X_{2}X_{5}$ &  \\
$X_{3}X_{6}$ &  \\
$Z_{2}Z_{3}Y_{4}Y_{6}$ &  $\bar{X}_{1} = X_{1}X_{2}X_{3}$ \\
$Z_{0}Z_{1}Z_{2}X_{3}Z_{4}Z_{5}$ &  $\bar{Z}_{1} = Z_{0}Z_{1}Z_{4}$ \\
\noalign{\smallskip}\hline\hline
\end{tabular}
\end{table}
It can be observed that the logical Pauli Z operator is already of Z-form with the lowest weight; Following the search algorithm,we now show in Fig~\ref{fig:figure1} a round-robin CZ circuit having been acted on three active qubits form a logical CZ gate for the bare [[7,1,3]] code.

\section{Resource Comparison and Simulation Scheme}
In this section, we first make a resource consumption comparison in terms of ancillary qubits consumed and multi-qubit gate counts between the 3-pieceable fault-tolerant CZ circuit and a 2-pieceable fault-tolerant CZ circuit constructed in previous research\cite{yoder2016universal}. Then, we give a numerical results to illustrate the performances of the two circuits. Finally, in order to improve the 2-pieceable fault tolerant CZ circuit, we design a new concatentation
\begin{figure}[tb]
\centering\includegraphics[width=0.7\linewidth]{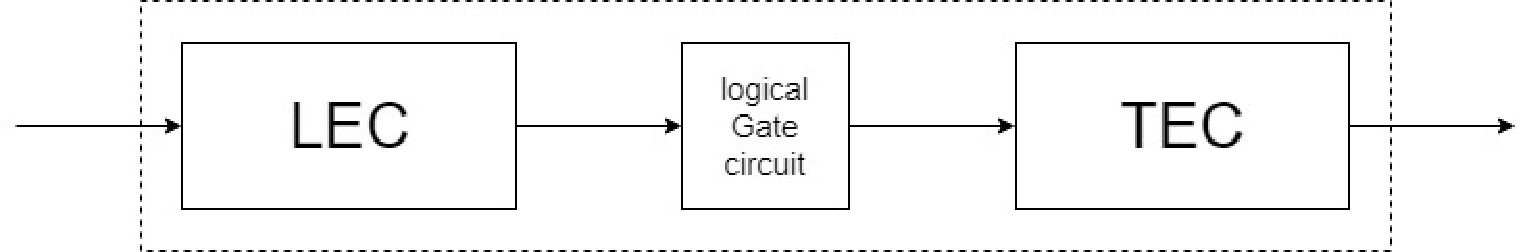}
\caption{\label{fig:figure2}Simulevel-1 Extended rectangle(1-exRec) consisting of leading and trailing error correcting circuits implementing the desired logical gate G}
\end{figure}
Following \cite{chamberland2017overhead}, we propose that it is contagious malignant error event that should be pay more attention, for it would cause more errors when we take the logical CZ circuit as the basic component for other high level concatenation circuits. So when performing Monte Carlo stabilizer simulations on the selected error subsets of the two circuits, we will take the contagious logical error rate as the major merit. Resource comparison of the two pieceable fault-tolerant circuits are conducted as in Table~\ref{tab:table3}
\begin{table}[tb]
\caption{\label{tab:table3}%
The resource comparison for implementing level-1 pieceable fault-tolerant logical CZ circuit on 5-qubit code and bare-7 qubit code. Note that the circuit used for comparison is an ideal variant, which only includes one cycle of syndrome measurements for each stabilizer. These circuits are included in whole exRec procedure, whereas in our simulation experiments.}
\begin{tabular}{l|l|l}
\hline \hline \noalign{\smallskip}
\textrm{Circuit} & \textrm{Ancillary} & \textrm{multiqubit gate}\\
\hline
5-qubit  & 196 & 457 \\
7-qubit  & 80  & 153 \\
Percent improvement & 59$\%$ & 66 $\%$ \\
\noalign{\smallskip}\hline\hline
\end{tabular}

\end{table}
\begin{figure}[tb]
\centering\includegraphics[width=0.9\linewidth]{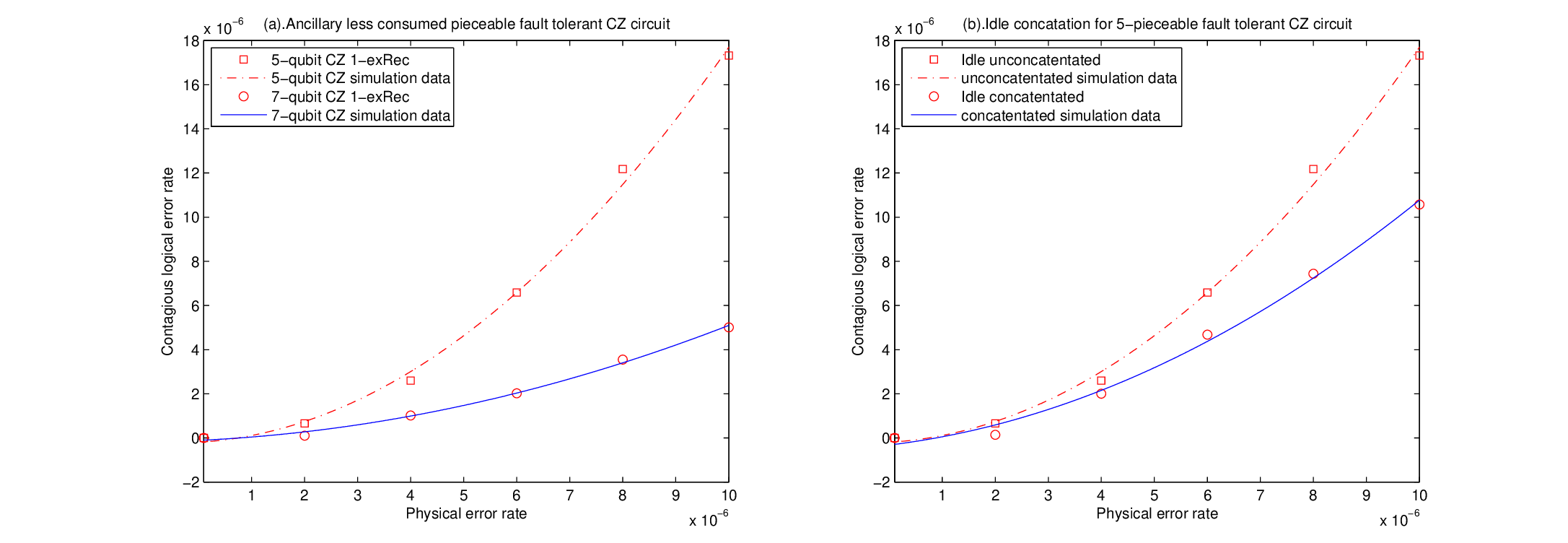}
\caption{\label{fig:figure3}Logical error rate for pieceable fault-tolerant CZ circuit on [[5,1,3]] code and bare [[7,1,3]] code under the standard depolarizing error model, with the error probability of a two-qubit gate and one-qubit gate being the same.}
\end{figure}
We propose that a lower logical error rate can be obtained for this 3-pieceable fault-tolerant logical CZ circuit, as its circuit volume is smaller than the one for [[5,1,3]] code, which makes obviously decrease the size of possible malignant error location sets. Here, we use the logical error rate calculation method presented in \cite{li2017fault}. Our simulation follows the 1-exREc formalism\cite{aliferis2005quantum} for level-1 encoding, this formalism is described in Fig~\ref{fig:figure2}; Before each round of simulation, we perform noise-free efficient encoding and obtain an encoded states.
Fig~\ref{fig:figure3} shows the logical error rates of the two circuits, with the assumption that the error probabilities of the one-qubit gate and two-qubit gate are the same.

\section{Summary and Conclusions}

In this paper, we presented a fault-tolerant control phase gate construction for a degenerate quantum code whose syndrome measurement circuit consumes at most two ancillary qubits per stabilizer. We also design an efficient non-uniform idle qubit concatenation scheme to further impress the possibly of propagated errors during the pieceable fault-tolerant logical CZ circuit.

In particular, for our chosen code, we believe that this approach to fault-tolerant quantum protocols can be used efficiently under standard depolarizing error models with relatively low logical error rates. In support of this, we  presented  Monte Carlo numerical simulation results.

As realistic error models become increasingly more relevant, the development of environment-specific fault-tolerant logical gates will become more important. In the meantime, the optimization of ancillary qubit resources and measurement times must be achieved for large-scale quantum computation. A natural direction of future work would be to study the pieceable fault-tolerant Toffli gate construction for given types of quantum code with high accuracy thresholds and utilize the optimization technique of circuit synthesis to further obtain  encoded quantum circuits with lower resource consumptions.

\section*{\sc Acknowledgements}
This work was supported by the National Natural Science Foundation of China under Grant No. 61572109

\end{document}